\documentclass[aps,prd,twocolumn,showpacs,superscriptaddress,nofootinbib]{revtex4-1}

\usepackage[utf8]{inputenc}
\usepackage{amssymb}
\usepackage{amsmath}
\usepackage{color}

\def\be{\begin{equation}}
\def\ee{\end{equation}}
\def\ba{\begin{align}}
\def\ea{\end{align}}

\begin{document}

\title{Superradiance without event horizons in General Relativity}

\author{Maur\'icio Richartz}
\email{mauricio.richartz@ufabc.edu.br}
\affiliation{Centro de Matem\'atica, Computa\c{c}\~ao e Cogni\c{c}\~ao, Universidade Federal do ABC (UFABC), 09210-170 Santo Andr\'e, SP, Brazil}
\author{Alberto Saa}
\email{asaa@ime.unicamp.br}
\affiliation{
Departamento de Matem\'atica Aplicada,
 UNICAMP, 13083-859 Campinas, SP, Brazil.}

\begin{abstract}
Superradiant scattering processes are studied in general relativistic systems which, unlike rotating and/or charged black holes, do not exhibit an event horizon. Inspired by Zel'dovich's seminal works on the amplification of waves by a rotating cylinder, we analyse, in the context of General Relativity, the possibility of superradiance for electromagnetic waves reflecting off a rotating star and for charged scalar perturbations impinging on a charged sphere. The role of energy dissipation in these systems is analysed and compared with the role of the event horizon in black hole superradiance.

\end{abstract}
\pacs{04.70.Bw, 04.40.Dg, 04.40.Nr, 03.50.De}
\maketitle

\section{Introduction}

A typical scattering process in nature consists in an incident wave interacting with an object (the scatterer) and being partly reflected and partly transmitted/absorbed. Usually, the amplitude of the reflected wave is smaller than the amplitude of the incident one, meaning that part of the energy carried by the incident wave is either absorbed by the scatterer or transmitted. However, in some peculiar setups (e.g.~rotating and/or charged systems), incident low frequency waves are able to extract energy from the scatterer, resulting in an amplified reflected wave. This kind of amplification process is known as superradiance~\cite{corinne,mauricio} and was first proposed and discussed by Zel'dovich, who analysed scalar~\cite{zeldovich1} and electromagnetic~\cite{zeldovich2,zeldovich3,bek_1} perturbations scattering off a rotating conducting cylinder in flat spacetime (herein referred to as Zel'dovich's cylinder). Soon after Zel'dovich's discovery, Misner~\cite{misner} investigated the possibility of superradiance in General Relativity. It was shown that incident (scalar, electromagnetic or gravitational) waves impinging on both charged~\cite{bek_2} and/or rotating~\cite{staro1,staro2} black holes can be superradiantly scattered if their frequency $\omega$ is sufficiently small, i.e.
\be 
0 < \omega < m\Omega_h + e\Phi_h, \label{superrad_cd}
\ee
where $m$ and $e$ are, respectively, the azimuthal number and the charge of the incident wave and $\Omega_h$ and $\Phi_h$ correspond, respectively, to the angular velocity and electric potential at the black hole horizon. More recently, it was shown that superradiant scattering is also possible in the context of analogue black holes~\cite{basak, basak2, mauricio2}.    

There are basically two crucial requirements for the occurrence of superradiance~\cite{mauricio}: first, there must be a region where negative energy modes, as defined by asymptotic observers, are allowed (or negative norm modes according to some inner product) and, second, there must exist a mechanism to extract these modes from the system. This mechanism is remarkably different when comparing superradiance in Zel'dovich's cylinder with black hole superradiance: in the first case, energy extraction occurs due to dissipation inside the cylinder; in the second, the event horizon acts as a one--way membrane, effectively preventing modes from exiting the black hole and reaching the asymptotic region. Mathematically, this difference translates into the conservation of the wave's stress energy tensor in black hole spacetimes, $\nabla_\mu T^{\mu \nu} = 0$, and non-conservation inside Zel'dovich's cylinder, $\nabla_\mu T^{\mu \nu} \neq 0$ (for scalar fields, this difference can be formulated equivalently in terms of the particle number current - see Sec.~\ref{sec3}). Another difference in superradiant systems concerns the origin of the negative energy modes as measured by asymptotic observers. For rotating black holes, the existence of such modes is usually associated with the ergosphere, i.e.~the region outside the event horizon where the asymptotic timelike Killing vector field $\mathbf{\partial_t}$ becomes spacelike. However, as has long been known, the existence of an ergosphere is not a necessary condition to the existence of negative energy modes~\cite{ruffini} (and, consequently, is not necessary to superradiance either). For instance, superradiance is possible in Zel'dovich's cylinder, which rotates in Minkowski spacetime, where no ergoregion is present. 

To the best of our knowledge about General Relativity, superradiant scattering has only been analysed before in black hole spacetimes. In view of that, our main objective in this work is to investigate the possibility of superradiance in general relativistic systems which do not posses an event horizon. We will, therefore, focus on general relativistic systems analogous to Zel'dovich's cylinder, in which negative energy modes are dissipated away instead of being trapped inside the event horizon of a black hole. 
Our first analysis concerns the scattering of electromagnetic waves impinging on a rotating conducting object which can serve as a model for a rotating star. We then discuss the possibility of superradiance for charged scalar fields scattered by a charged sphere whose exterior metric is the Reissner-Nordstr\"om metric. Besides the absence of event horizons, in order to avoid the occurrence of the so-called ergosphere instability~\cite{insta1, insta2, insta3, insta4}, we also assume no ergospheres in our spacetimes.

\section{Superradiance in rotating stars}

Inspired by Zel'dovich's original work~\cite{zeldovich1} concerning the possibility of superradiance in dissipative systems, we shall consider the scattering of electromagnetic perturbations impinging on a general relativistic rotating star. The spacetime is assumed to be stationary and axisymmetric, with corresponding timelike and spacelike Killing vector fields given, respectively, by $\xi^\mu _t = (\partial_t)^\mu$ and $\xi^\mu _{\phi} = (\partial_\phi)^\mu$. In spherical coordinates, the most general metric $g_{\mu \nu}$ representing such spacetime can be written as~\cite{bardeen, chandra} 
\begin{align} 
ds^2 = -e^{ 2\nu}dt^2 + e^{2\mu} dr^2 + e^{2\lambda} r^2 d\theta ^2 \nonumber \\
 + r^2 \sin ^2 \theta e^{2 \alpha} \left( d\phi - \varpi dt \right)^2, \label{metric_x}
\end{align}
where $\mu$, $\nu$, $\lambda$, $\alpha$ and $\varpi$ are functions of $r$ and $\theta$ only. The four-velocity of the star is $u^{\mu}=\gamma \left(1,0,0,\Omega\right)$, where $\Omega=\Omega(r,\theta)$ is the angular velocity of the star and $\gamma$ is given by
\be
\gamma = \left[e^{2\nu} - e^{2 \alpha} r^2 \sin ^2 \theta  \left( \varpi - \Omega \right)^2 \right]^{-\frac{1}{2}}.
\ee
Note that there is a degree of freedom unspecified in \eqref{metric_x} between the coordinates $r$ and $\theta$: usually one sets $\alpha = \lambda$ to study perturbations of slowly rotating stars~\cite{hartle1} or $\lambda = \mu$ to study rapidly rotating stars~\cite{hartle2} -- in our paper, we leave this degree of freedom unspecified.  

The star is assumed to have a non-zero conductivity $\sigma$ and, possibly, a non-zero charge density $\rho$ near its surface. Outside the star (vacuum), we have $\sigma=\rho=0$. We further assume that there are no event horizons in our spacetime and that $\xi^{\mu}_t=(\partial_t)^{\mu}$ is always timelike (therefore, no ergospheres are present). See Refs.~\cite{luciano,rezzolla2} for astrophysical applications of similar models of rotating stars.  

  Let us now investigate the scattering of electromagnetic waves impinging on the rotating star metric described above. The evolution of such perturbations is determined by Maxwell's equations,
\begin{align} 
\nabla _ \mu{}^*\!F^{\mu \nu} = 0 \quad \left(\text{equivalently, }\nabla_{[\mu}F_{\nu \sigma]} = 0 \right), \label{max_eqn1} \\
\nabla _ \nu F^{\mu \nu} = 4 \pi J^{\mu} = 4\pi \left[\sigma F^{\mu \nu}u_{\nu} + \rho u^{\mu} \right] , \label{max_eqn2}
\end{align} 
where $F_{\mu \nu}$ and ${}^*\!F_{\mu \nu}$ are, respectively, the Faraday tensor and its dual, and $J^{\mu} = \sigma F^{\mu \nu}u_{\nu} + \rho u^{\nu}$ is the total electromagnetic current. From the axisymmetry and the stationarity of our system, we conclude that the $t$ and $\phi$ dependence of $F_{\mu \nu}$ is given by  $e^{-i\omega t} e^{im\phi}$. Additionally, the electromagnetic stress energy tensor associated with $F_{\mu \nu}$ is given by
\be \label{tuv}
T_{\mu \nu} = \frac{1}{4 \pi} \left[F_{\mu \sigma}F_{\nu}{}^{\sigma} - \frac{1}{4}g_{\mu \nu}F^{\sigma \rho}F_{\sigma \rho} \right].
\ee
  
  Neglecting the backreaction of these electromagnetic perturbations on the background spacetime, we can assume the metric~\eqref{metric_x} to be fixed. 
As explained in the introduction, the occurrence of superradiance in systems without event horizons is related to dissipative effects, i.e.~the non-conservation of the stress energy tensor. In fact, for electromagnetic perturbations we have $\nabla_{\mu}T^{\mu \nu} = J_{\sigma}F^{\sigma \nu}$, which is non-zero inside the star. Consequently, the associated energy current $\widetilde{J}_\mu = - T_{\mu \nu} \xi^{\nu}_t$ is also not conserved inside the star,
\be \label{divtuv}
\nabla^{\mu} \tilde{J}_\mu = J^\mu F_{\nu \mu} \xi^\nu_t =  \left( \sigma F^{\mu \sigma} u_{\sigma} + \rho u^{\mu} \right)  F_{\nu \mu} \xi^\nu_t.
\ee
Using Maxwell's equations \eqref{max_eqn1}, one can show that
\be
\partial_\mu F_{30} = i\omega F_{\mu 3} - imF_{0 \mu},
\ee
which, after substitution into~\eqref{divtuv}, together with the expressions for $u^\mu$ and $\xi_t^{\mu}$ imply that 
\be 
\nabla^{\mu} \tilde{J}_\mu =  \gamma \rho \Omega F_{03} - \gamma \frac{\sigma}{\omega} F^{\mu}{}_0 \left[ \left(\omega - m \Omega \right)  F_{\mu 0} - i \Omega \partial_ \mu F_{30}\right] . \label{divju}
\ee
Now, in order to calculate the energy flux at infinity we would like to integrate the expression above over the region K of the spacetime bounded by constant time hypersurfaces at $t_1$ and $t_2$ and by constant radius hypersurfaces at $r_1 \rightarrow 0$ and $r_2 \rightarrow \infty$. Using Gauss' law, we obtain
\be 
\int_{K} \! \! \! \nabla^{\mu} \widetilde{J}_\mu  dV = 
 \oint_{\partial K} \! \! \! \! \! \! \widetilde{J}_\mu n^{\mu} d\Sigma, \label{gauss1}
\ee
where $dV=\sqrt{-g}d^4x$ is the 4-volume element, $d\Sigma$ is a 3-volume surface element and $n^{\mu}$ is the outward unit normal to $\partial K$, the boundary of $K$. Because of the stationarity of the system, the integrals over the constant time hypersurfaces cancel each other and the RHS of equation~\eqref{gauss1} can be rewritten as
\begin{align}
-\lim_{r_2 \to +\infty} \oint_{r=r_2} \! \! \! \! \! \! T_{01} e^{-\mu} d\Sigma 
 + \lim_{r_1 \to 0} \oint_{r=r_1}  \! \! \! \! \! \! T_{01} e^{-\mu} d\Sigma, \label{wide} 
\end{align}
where we have used the fact that $n^{\mu} = \pm \left(0,e^{-\mu},0,0\right)$ for the constant radius hypersurfaces.

The first and second terms in the expression above can be interpreted, respectively, as the energy fluxes through through infinity and through the origin. Since there is no singularity at $r=0$, the energy flux through the origin vanishes (in principle, this can be demonstrated by modelling the matter inside the star, solving Maxwell's equations in the limit $r\rightarrow 0$ and requiring regularity of the fields). Combining equations \eqref{divju}, \eqref{gauss1} and \eqref{wide}, and using the fact that $J^{\mu}$ vanishes outside the star (because $\sigma=\rho=0$), one obtains  
\begin{widetext}
\be
\int_{K_1 \cup K_2} \! \! \! \! \! \! \nabla^{\mu} \widetilde{J}_\mu  dV = \int_{K_1}  \! \! \! \gamma \left\{ \rho \Omega F_{03} - \frac{\sigma}{\omega} F^{\mu}{}_0 \left[ \left(\omega - m \Omega \right)  F_{\mu 0} - i \Omega \partial_ \mu F_{30}\right] \right\}  dV = 
-\lim_{r_2 \to +\infty} \oint_{r=r_2} \! \! \! \! \! \! T_{01} e^{-\mu} d\Sigma. \label{wide3} 
\ee
\end{widetext}

When deriving superradiance in Zel'dovich's cylinder~\cite{zeldovich2, zeldovich3, bek_1}, the usual procedure is to restrict the analysis to certain polarization modes. Similarly, in the rotating star case, we now restrict our analysis to modes which have vanishing $F_{03}$. Physically, this means that zero angular momentum observers (ZAMOs)~\cite{bardeen} will measure a vanishing electric field in the $\hat \phi$ direction. For these modes, eq.~\eqref{wide3} reduces to             
\be
\lim_{r_2 \to +\infty} \oint_{r=r_2} \! \! \! \! \! \! T_{01} e^{-\mu} d\Sigma =  \frac{ \sigma}{\omega} \int_{K_1}  \! \! \! \left(\omega - m \Omega \right) \gamma  F^{\mu}{}_0 F_{\mu0}  dV   
 \label{flux_infty2}.
\ee

The LHS of the expression above corresponds to the energy flux entering the system from $+\infty$ and is always negative if $0< \omega < m\Omega_{min}$, where $\Omega_{min}$ denotes the minimum value of the angular velocity of the star. In other words, if the frequency is sufficiently small, the flux of energy being radiated away from the star is larger than the incident energy flux, characterizing superradiance. Furthermore, it is also possible to calculate the flux of angular momentum by replacing the energy current $\widetilde{J_{\mu}}$ with the angular momentum current $T_{\mu \nu} \xi^{\nu}_{\phi}$ in the calculations above. Proceeding this way, one is able to show that not only energy, but also angular momentum is extracted from the star. One can, therefore, conclude that low-frequency electromagnetic waves will be superradiantly scattered by the star, extracting (part of) its rotational energy in the process. 

We end this section by comparing our analysis for rotating stars with the well-known results for rotating black holes. Due to the no-hair theorem, the only black hole parameters that can interfere in the superradiant amplification are its mass, charge and angular momentum. In general, larger angular velocities (i.e.~larger angular momenta) will produce larger energy fluxes - note, however, that for a given mass there is a limit to how rapidly a black hole can rotate before becoming a naked singularity. On the other hand, for a rotating star, besides a similar dependence on the angular velocity, the energy flux will also depend on the conductivity $\sigma$. There is no limit on how large $\sigma$ can be (the usual MHD assumption is $\sigma \rightarrow \infty$) and, therefore, a superficial analysis leads to the conclusion that the energy flux, being directly proportional to $\sigma$, can be arbitrarily large for rotating stars. However, a more detailed analysis is required since larger conductivities will generally lead to smaller electromagnetic fields, which could, in principle, compensate for the increase of $\sigma$ in~\eqref{flux_infty2}.
 
\section{Superradiance in charged spheres} \label{sec3}

In blackhole spacetimes it is known that superradiance is possible in both rotating (Kerr metric) and charged (Reissner-Nordstr\"om metric) situations. As we shall see below, a similar statement concerning dissipative superradiance is also true. In the previous section, we have shown the possibility of superradiance in a rotating system without an event horizon. Now, in our second example, we shall consider the scattering of a charged scalar field impinging on a general relativistic sphere of mass $M$, charge $Q$ and radius $r_0$. The corresponding spacetime metric is given by
\be \label{gen_metric}
ds^2 = -e^{2\nu(r)} dt^2 + e^{2\mu(r)} dr^2 + r^2 d\theta^2 + r^2 \sin^2 \theta d\phi ^2,
\ee
where $\left(t,r,\theta,\phi\right)$ are the usual spherical coordinates and $\mu$ and $\nu$ are functions of $r$ only. Outside the sphere ($r > r_0$), the vacuum Einstein-Maxwell equations are satisfied and, therefore, the spacetime is described by the RN metric,
\be
e^{2\nu(r)}= e^{-2\mu(r)} = \frac{\Delta}{r^2} = \frac{r^2 - 2Mr + Q^2}{r^2}. \label{reis}
\ee 
Since we do not want to consider black hole spacetimes, we are implicitly assuming that $r_0$ is larger than the radius of the RN black hole, i.e. $r_0> M + \sqrt{M^2 - Q^2}$. There are no ergospheres in our spacetime either since we assume that $\xi_t ^{\mu} = \mathbf{\partial_t}^\mu$ is always timelike. 

The interior metric ($r<r_0$), on the other hand, is determined by solving the non-vacuum Einstein-Maxwell equations. Because of the spherical symmetry, the only non-zero component of the Faraday tensor $F_{\mu \nu}$ is 
\be \label{potential}
F_{01} = -F_{10} = \frac{d\Phi}{dr}  = - \frac{q}{r^2} e^{\nu + \mu},
\ee
where $\Phi(r)$ is the associated electric potential and $q(r)$ represents the charge inside a sphere of radius $r$. The corresponding electromagnetic 4-potential is given by $A_{\mu} = \left(-\Phi(r),0,0,0\right)$. We refer the reader to Ref.~\cite{ivanov} for a comprehensive review of interior RN solutions.

Having described the background spacetime, we now proceed to investigate the possibility of superradiance by a massless scalar field $\psi$ with electric charge $e$, which is assumed to be minimally coupled to the electric potential $A_{\mu}$. The propagation of this scalar field is governed by the Klein--Gordon (KG) equation,
\be \label{kg}
 \left( \nabla^\mu - ie A^{\mu} \right)\left( \nabla_\mu - ie A_{\mu} \right)  \, \psi = 0,
\ee
which can be derived from the following action,
\be \label{action1}
S = \int  \left[ \left( \nabla_\mu - ie A_{\mu} \right) \psi \right] \left[ \left( \nabla^\mu + ie A^{\mu} \right) \psi ^* \right] \sqrt{-g} \, d^4x.
\ee
Dissipative effects in this system can be included by introducing a new term in the KG equation. More precisely, we modify equation \eqref{kg} in the reference frame of an stationary observer with respect to the charged star according to
\be \label{kg2}
 \left( \nabla^\mu - ie A^{\mu} \right)\left( \nabla_\mu - ie A_{\mu} \right)  \, \psi - a\left(\partial_t - i e A_t \right)\psi = 0,
\ee
where $a$ is a constant representing the absorptive properties of the medium. This extra term is completely analogous to the dissipative term considered by Zel'dovich in his original analysis~\cite{zeldovich1}. 

By introducing the ansatz $\psi = R(r) Y_{\ell m} (\theta, \phi) e^{- i \omega t}$, the dissipative KG equation above becomes separable. The angular functions $Y_{\ell m}(\theta, \phi)$ are the spherical harmonics of orbital number $\ell$ and azimuthal number $m$, $\omega$ is the frequency of the wave and the radial function $R$ satisfies the following equation, 
\begin{align} 
e^{-\nu - \mu}\frac{d}{dr}\left( e^{\nu - \mu}r^2 \frac{dR}{dr}  \right) + \left[e^{- 2\nu} r^2 \left(\omega - e \Phi(r) \right)^2 \right. \nonumber \\ 
\left. +i a r^2 \left( \omega - e \Phi(r) \right) - \ell(\ell + 1) \right]R = 0. \label{rad_eqn_2}
\end{align}

The full solution of the equation above depends on the specifics of both the background metric and the electric potential. In our superradiance analysis it is sufficient to know the explicit solution only in the asymptotic limit, 
\be \label{sol_far2}
R(r) = \frac{1}{r} e^{- i \omega r} + \frac{\mathcal{R}_{\omega \ell}}{r} e^{+ i \omega r}, \qquad    r \rightarrow \infty, 
\ee
where $\frac{1}{r} e^{- i \omega r}$ represents the incident wave and $\frac{1}{r}e^{+ i \omega r}$ represents the reflected wave with amplitude $R_{\omega \ell}$.

A possible way to determine whether superradiance occurs or not is to calculate the so-called particle number current $\widetilde{J}_{\mu}$,
\be 
\widetilde{J}_{\mu} = i \left[ \psi \left( \nabla _\mu + i e A_{\mu} \right) \psi^* - \psi^* \left( \nabla _\mu - i e A_{\mu} \right) \psi \right], \label{jtilde}
\ee
associated with both incident and reflected waves. This quantity, being the Noether-current associated with the symmetry $\psi \rightarrow  e^{i \beta} \psi$ of the action \eqref{action1}, is conserved for solutions $\psi$ of the original KG equation~\eqref{kg}, i.e.~$\nabla_\mu \widetilde{J}^\mu = 0$; for the dissipative KG equation~\eqref{kg2}, however, the same does not happen and
\begin{align}
\nabla_\mu \widetilde{J}^\mu = -i a \left( \psi^* \partial_t \psi - \psi \partial_t \psi ^* \right) - 2 e a A_t \vert \psi \vert ^2 \nonumber \\
 = -2a (\omega - e \Phi(r))\vert \psi \vert ^2,
\end{align} 
where we have used the explicit time-dependence of the scalar field and the fact that $A_t=-\Phi(r)$ to obtain the last equality. Integrating the above expression over the same region K described after equation~\eqref{divju} and using Gauss' law, one obtains 
\begin{align} 
\int_{K}  \nabla_\mu \widetilde{J}^\mu dV = \int_{K} (-2a) \left[\omega - e \Phi(r)\right]\vert \psi \vert ^2 dV \nonumber \\ = \oint_{\partial K} \widetilde{J}_\mu n^{\mu} d\Sigma, \label{gausss1}
\end{align}
where, again, $dV=\sqrt{-g}d^4x$ is the 4-volume element, $d\Sigma$ is a 3-volume surface element and $n^{\mu}$ is the outward unit normal to $\partial K$, the boudary of $K$. Following the same procedure used in the rotating case, we can use equation~\eqref{gausss1} to obtain the flux through infinity,
\begin{widetext}
\be
\lim_{r_2 \to +\infty} \oint_{r=r_2} \! \! \! \!  \tilde{J}_\mu n^{\mu} d\Sigma = -8\pi a\Delta t \int\limits_{0}^{r_0} \left[\omega - e\Phi(r) \right] e^{\mu + \nu} r^2 \vert R \vert^2 dr
\ge -8 \pi a \Delta t \left[\omega - e\Phi(r_0) \right] \int\limits_{0}^{r_0} e^{\mu + \nu} r^2 \vert R \vert^2 dr, \label{finfty2}
\ee
\end{widetext}
where $\Delta t = t_2 - t_1$. In order to write the expression above, we have performed the temporal and angular integrals and have used the fact that the electric potential is a decreasing function of $r$ [see eq.~\eqref{potential}]. Consequently, if the frequency of the incident wave is sufficiently small, $0 < \omega < e \Phi(r_0)$, we conclude that the particle number flux through infinity is negative and, therefore, the flux associated with the reflected wave is larger than the one associated with the incident wave, characterizing superradiance.
    
It is also possible to derive an expression for the reflection coefficient $\vert R_{\omega \ell} \vert ^2$. By rewriting the LHS of \eqref{finfty2} with the help of eqs.~\eqref{sol_far2} and~\eqref{jtilde}, we obtain 
\begin{align}
\left| R_{\omega \ell} \right|^2 = 1 - \frac{a}{\omega} \int_{0}^{r_0} \left(\omega - e \Phi(r)\right) e^{ \mu + \nu} r^2 \left| R(r) \right|^2 dr \nonumber \\
> 1 - \frac{a(\omega - e \Phi(r_0))}{\omega} \int_{0}^{r_0} e^{ \mu + \nu} r^2 \left| R(r) \right|^2 dr. \label{reflec}
\end{align}
Therefore, for sufficiently low frequencies $0 < \omega < e \Phi(r_0)$, the reflection coefficient is greater than one. Note that the role of the absorption parameter $a$ is analogous to the role of the conductivity $\sigma$ in the rotating star case. Like $\sigma$, the absorption parameter $a$ is directly proportional to the superradiant amplification. Furthermore, an increase in $a$ does not necessarily mean a larger reflection coefficient since the radial function $R(r)$ appearing in~\eqref{reflec} also depends on $a$ [see~\eqref{rad_eqn_2}]. Finally, we note that for the original, non-dissipative ($a=0$) KG equation, we have $\left| R_{\omega \ell} \right|^2 = 1$, meaning that the incoming particle number flux equals the reflected one, i.e. no absorption/amplification takes place, as expected. 

\section{Final Remarks}

Motivated by Zel'dovich's works on superradiance~\cite{zeldovich1, zeldovich2}, we have discussed in this paper the possibility of superradiant scattering in general relativistic systems without event horizons. 
However, instead of the cylindrical scatterer of Zel'dovich, we have considered a rotating star and a charged sphere in our analysis. In fact, we have shown that electromagnetic waves incident on a rotating star and charged scalar perturbations impinging on a charged sphere can be amplified if their frequencies are sufficiently low. Similarly to Zel'dovich's cylinder, this type of superradiant scattering is possible only because of dissipation of negative energies inside the scatterer (compare with black hole superradiance, where, instead of energy dissipation, there is an event horizon present). We believe that, since Zel'dovich's cylinder rotates in a flat spacetime, our analysis is the first one of this kind of dissipative superradiance in General Relativity.  

We also note an interesting difference between the two systems analysed in this paper: for the charged sphere, we have derived superradiance using a Noether current of the non-dissipative KG equation; for the rotating star, instead, we have used the stress energy tensor of the Maxwell field and the associated energy current. However, we believe that it is also possible to derive superradiance using the KG stress energy tensor in the charged sphere case and a Noether current in the rotating star case (see Ref.~\cite{wald} for a similar discussion concerning black hole superradiance). 

Finally, we also point out a few characteristic details of our derivation of superradiance in the rotating star case. First, we have assumed that the permittivity $\epsilon$ and the permeability $\mu$ of the spacetime do not change significantly between the inside and the outside of the star. Taking into account possible differences in $\epsilon$ and $\mu$ would probably enhance the superradiant amplification -- see Ref.~\cite{bek_1} for a detailed analysis in Zeldovich's cylinder. Furthermore, our derivation did not depend on any kind of separability of Maxwell's equations in the variables $r$ and $\theta$. The derivation of superradiance for a Kerr black hole, on the other hand, usually relies on the separability of the Newman-Penrose scalars $\phi_0$, $\phi_1$, $\phi_2$ and the associated Teukolsky equations~\cite{teuko1, teuko3}.  

\begin{acknowledgments}
The authors are grateful to FAPESP and CNPq for the financial support.
\end{acknowledgments}

\bibliographystyle{apsrev4-1}
%

\end{document}